\documentstyle[12pt,epsfig]{article} 
\newlength{\dinwidth}                       
\newlength{\dinmargin}                      
\newcommand{\pmb}[1]{%
        \setbox0=\hbox{#1}%
        \kern-.02em\copy0\kern-\wd0
        \kern+.04em\copy0\kern-\wd0
        \kern-.02em\raise.0217em\box0}
\newcommand{\vek}[1]{
        \mathchoice{\mbox{\boldmath$#1$}}%
        {\mbox{\boldmath$#1$}}%
        {\pmb{$\scriptstyle#1$}}%
        {\pmb{$\scriptscriptstyle#1$}}}
\newcommand{\lsim}{
 \mathrel{\setbox0=\hbox{$<$}\raise0.6ex\copy0\kern-\wd0
 \lower0.65ex\hbox{$\sim$}}}
\newcommand{\gsim}{
 \mathrel{\setbox0=\hbox{$>$}\raise0.6ex\copy0\kern-\wd0
 \lower0.65ex\hbox{$\sim$}}}

\begin{document}
\leftline{\hfill TUM/T39-96-18, OSU-96-0701}
\vspace*{1cm}
\begin{center}  \begin{Large} \begin{bf}
Coherent $\rho$ Production from Polarized Deuterium\footnote{Work 
supported in part by grants from BMBF, GIF and the DoE.}\\
  \end{bf}  \end{Large}
  \vspace*{5mm}
  \begin{large}
L. Frankfurt$^{ae}$, W. Koepf$^{b}$, 
J. Mutzbauer$^{c}$, \\ 
\underline {G. Piller}$^{c}$, 
M. Sargsyan$^{af}$, M. Strikman$^{de}$
  \end{large}
\end{center}
$^a$ School of Physics and Astronomy, Tel Aviv University,
     Tel Aviv   69978, Israel\\
$^b$ Department of Physics, Ohio State University, Columbus, OH 43210, USA\\
$^c$ Physik Department, Technische Universit\"{a}t M\"{u}nchen,
      D-85747 Garching, Germany \\
$^d$ Department of Physics, Pennsylvania State University, 
     University Park, PA  16802, USA\\
$^e$ Institute for Nuclear Physics, St. Petersburg, Russia\\
${^f}$ Yerevan Physics Institute, Yerevan 375036, Armenia\\
\begin{quotation}
\noindent
{\bf Abstract:}
We discuss the coherent  leptoproduction of vector mesons 
{}from polarized deuterium as a tool to investigate the evolution of 
small size quark-gluon configurations. 
Kinematic regions are determined where the final state interaction 
of the initially produced quark-gluon wave packet contributes 
dominantly to  the production cross section.
Two methods for an investigation of color coherence effects, 
which are appropriate for future experiments at HERMES, are suggested.
\end{quotation}
\section{Scales in color coherence}
High energy exclusive production processes from nucleon targets 
are determined by the transition of initial partonic wave 
functions  to final hadronic states. 
Interesting details about such transition amplitudes can be 
obtained by embedding the production process into nuclei,  
where the formation of a particular final state hadron is probed 
interactively via the interaction with 
spectator nucleons (for a review see \cite{FMS}).

In this context, we discuss the coherent photo- and leptoproduction 
of vector mesons from polarized deuterium at large photon energies 
$\nu \gsim 4\,GeV$:
\begin{equation}
\gamma^* + \vec d \longrightarrow V + d.
\end{equation}
The corresponding amplitude can be split into two pieces:
In the single scattering term only one nucleon participates 
in the interaction. 
This is in contrast to the double scattering contribution.  
Here 
the (virtual) photon interacts with one of the nucleons 
inside the target and produces an intermediate hadronic state   
which subsequently re-scatters from the second nucleon 
before forming the final state vector meson. 
At small $Q^2 \lsim 1\,GeV^2$ exclusive vector meson production 
{}from deuterium is well understood in terms of vector meson dominance. 
In this framework the final state vector meson is formed  
instantaneously in the interaction of the photon with  
one of the nucleons from  the target \cite{BSJ}. 
On the other hand, in the limit of large $Q^2 \gg 1\,GeV^2$ 
perturbative QCD calculations show that photon-nucleon scattering  
yields rather a small size, color singlet quark-gluon wave 
packet (ejectile) than a soft vector meson \cite{BFGMSS}. 
At high energies $\nu$ the final state interaction of such an ejectile 
with the second nucleon  should differ substantially from the 
final state interaction of a soft vector meson. 
Therefore the magnitude of the final state interaction contains 
interesting informations about the initially produced ejectile and 
its evolution while penetrating through the target. 

The ejectile wave packet and its propagation are  characterized 
to a large extent  by the following scales: 
The average {\bf transverse size} of the ejectile wave packet, 
$b_{ej}$, which for the case of longitudinal photons is
$\approx  4 \ldots 5/Q$ for $Q^2 \geq 5 GeV^2$ for the contribution 
of the minimal Fock space component \cite{FKS}.
For these $Q^2$  it amounts to less than 
a quarter of the typical diameter of a $\rho$-meson ($\approx 1.4\,fm$).
Furthermore  the initially produced small quark-gluon wave packet does 
not, in general, represent an eigenstate of the strong interaction 
Hamiltonian.
Expanding the ejectile in hadronic eigenstates one 
finds  that  inside a nuclear target all hadronic components, except 
the measured vector meson, are filtered out via final state interaction 
after a typical {\bf formation time}:  
$\tau_f \approx 2\nu/\delta m_V^2$. 
Here $\delta m_V^2$ is a characteristic squared mass difference 
between low-lying vector meson states, which is related to
the inverse slope of the corresponding Regge trajectory 
($\delta m_V^2 \sim 1 GeV^2$).
If the formation time is larger than the nuclear radius, the 
ejectile will resemble an approximate eigenstate of the strong interaction 
while penetrating through the nuclear target.
Therefore final state interaction  should decrease with rising 
$Q^2$ at large photon energies. This phenomenon is usually called 
color coherence.  

However  the coherent vector meson production cross section, 
including its contribution from double scattering,   
is sensitive also to the {\bf coherence length}:  
$\lambda \approx 2\nu/(m_V^2 + Q^2)$. 
The latter characterizes  the minimal longitudinal momentum transfer 
$k_L \approx \lambda^{-1}$  required for the coherent production 
of the vector meson. (Here we omit the $t$-dependence of 
$\lambda$ and $k_L$ which is  discussed in details in ref.\cite{FKMPSS}.)
Since the deuteron has to stay intact,  one finds dominant contributions 
to the production cross section for $\lambda > R_d$, 
where $R_d\approx 4\,fm$ is the deuteron radius.  
Consequently  vector meson production amplitudes from 
nucleons at a similar impact parameter but at different 
longitudinal positions will interfere and add up coherently.
A decrease  of the coherence length leads to a decrease of the 
coherent vector meson production cross section. 
Therefore
if one investigates the double scattering contribution 
in different kinematic regions  one should be careful in interpreting
a variation of the final state interaction as 
a modification of the the ejectile wave function. 
First possible effects arising from a change in the coherence length 
have to be accounted for.

\section{Single versus double scattering}
In the single scattering contribution the vector meson 
is produced off one of the nucleons in the target, while 
the second nucleon does not participate in the interaction.
The corresponding Born amplitude is determined by the vector meson production 
amplitude $f^{\gamma^* p(n) \rightarrow V p(n)}$ from the proton or 
neutron, respectively, and the deuteron form factor $S_d^j$
(where $j$ indicates the dependence of the form factor on 
the target polarization):
\begin{equation} \label{eq:single}
F^{(1)} = f^{\gamma^* p \rightarrow V p}(\vek k_{\perp}) 
          S_d^j(-\vek k_{\perp}/2,k_{\_}/2)  + 
          f^{\gamma^* n \rightarrow V n} (\vek k_{\perp})  
          S_d^j(\vek k_{\perp}/2,-k_{\_}/2). 
\end{equation}
The presence of $k_{\_} = k_0 - k_L$ in the form factors accounts for 
the recoil of the deuteron.  
If the target polarization is chosen perpendicular to the 
momentum transfer  $\vek k$, 
at large $\nu$ 
only the difference of the monopole 
and quadrupole form factor, $S_d^j = S_d^0 - S_d^2$, enters 
in (\ref{eq:single}). 
In terms of the $S$- and $D$-components of 
the deuteron wave function  these are  
$S_d^0(k) = \int_0^{\infty}dr\, j_0(kr)[u^2(r) + w^2(r)]$ 
and 
$S_d^2(k) = \sqrt{2} \int_0^{\infty}dr\, j_2(kr)w(r)[u(r) - 
 w(r)/\sqrt{8}]$. 
It is important to realize that the monopole and quadrupole 
form factor are  equal at $k = |\vek k| \approx 0.35\,GeV$.
This generates a zero in $S_d^0 - S_d^2$ and consequently a 
node in the single scattering  contribution to the 
vector meson production cross section at 
$t = t_d \approx - \vek k^2 \approx - 0.5\,GeV^2$ \cite{FKMPSS}
(see also the discussion of elastic hadron-deuteron scattering in
ref.\cite{FG}). 
It should be emphasized that the latter is determined solely 
by the deuteron wave function and does not depend  
on details of the nucleon production amplitude,
$f^{\gamma^* p(n) \rightarrow V p(n)}$. 
Thus  we have identified a kinematic window where the 
single scattering contribution vanishes and the double scattering 
contribution  can be investigated to high accuracy.
A similar behavior of the single scattering amplitude can be 
achieved for a deuteron polarization along either
$\hat {\vek \kappa}  = (2 \vek q+\vek k)/|2 \vek q+\vek k|$
or $\hat {\vek n} = \hat {\vek k} \times \hat {\vek \kappa}$, 
where $\vek q$ stands for the photon  three-momentum. 
On the other hand,
for an unpolarized deuteron target the sum of the monopole and 
quadrupole form factor  enters in the Born amplitude. 
As a consequence  the single scattering contribution always 
dominates the vector meson production cross section 
at moderate $ - t \lsim 1\,GeV^2$, leaving no favorable 
kinematic window for an investigation of final state interaction.

The double scattering amplitude stems from the final 
state interaction of an initially produced hadronic state. 
Expanding the latter  in hadronic eigenstates $h$ yields:
\begin{equation} 
F^{(2)} \approx \frac{i}{2} \sum_h \int \frac{d^2 \vek k_{\perp}'}{(2 \pi)^2}
S_d^j(\vek k_{\perp}',k_{\_}/2) 
f^{\gamma^* p \rightarrow h p}(\vek k_{\perp}/2 - \vek k_{\perp}') 
f^{h n \rightarrow V n} (\vek k_{\perp}/2 + \vek k_{\perp}') 
+ (p \leftrightarrow n). 
\end{equation}
The transfered momentum is split between both interacting nucleons. 
Therefore  if the re-scattering amplitude of the ejectile 
$f^{h n(p) \rightarrow V n(p)}$ is sizable, double scattering will 
be important in the region of moderate and large $-t\gsim 0.4 \,GeV^2$.

In Fig.1 we show the differential cross section for the coherent 
$\rho$-production from 
deuterium  polarized perpendicular to the momentum transfer, 
\begin{equation}
\frac{d\sigma_d}{dt} =   \frac{1}{16\pi} 
\left( |F^{(1)}|^2 + 2 Re \left(F^{(1)*} F^{(2)}\right) 
+ |F^{(2)}|^2\right),
\end{equation}
calculated within vector meson dominance \cite{FKMPSS}.
In this framework  the double scattering amplitude accounts for the 
sizable re-scattering of the soft $\rho$-meson. 
The energy and momentum transfer are taken within the 
kinematic domain of the HERMES experiment.
\begin{figure}
\centering{\ \epsfig{figure=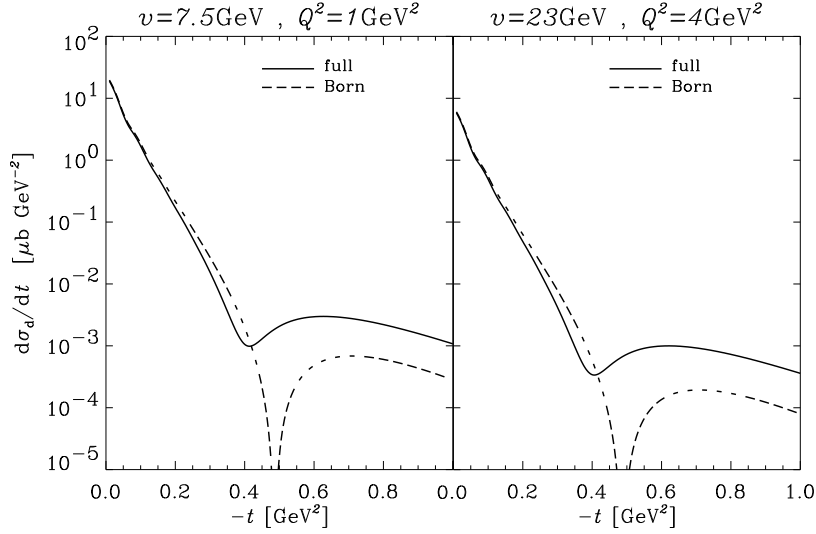,height=7cm}}
\caption {\it The differential production cross section 
         for coherent $\rho$-production from 
         polarized deuterium for different values of $Q^2$ and $\nu$. 
         The target polarization is chosen perpendicular to the momentum 
         transfer. The full line is the result for the production cross 
         section within vector meson dominance.
         The dashed curve shows the Born contribution. 
}
\end{figure}
Figure 1 demonstrates that the minimum of the single scattering amplitude 
(at $-t \approx 0.5\,GeV^2$)
is completely filled by final state interaction. 
Furthermore  double scattering exceeds the Born contribution 
by a factor $\sim 5$  at $-t > 0.5\,GeV^2$.

\section{Signatures for color coherence}       

Of course  vector meson dominance is appropriate for small 
$Q^2 \lsim 1\,GeV^2$ only. 
However  we want to study  the dependence of the 
final state interaction on $Q^2$ and $\nu$ 
to obtain informations on the ejectile wave function 
and its propagation through the nucleus. 
For this purpose we suggest investigations which are to a large 
extent independent on  details of the ejectile production 
amplitude.
    
In this respect  it is most promising to study coherent 
vector meson production from deuterium polarized perpendicular  
to the momentum transfer $\hat {\vek k}$  
(or parallel to $\hat {\vek \kappa}$ or $\hat {\vek n}$)  
at $t \sim t_d$. 
Here  the single scattering contribution to the cross section 
is very small  and even vanishes at $t = t_d$ as discussed above.
Exploring the production cross section in this kinematic window 
for different $Q^2$ and $\nu$
yields direct informations on the ejectile wave function. 
Since the transverse size of the ejectile shrinks with rising 
$Q^2$, perturbative QCD suggests that the vector meson production 
cross section should decrease and ultimately vanish at $t = t_d$ for  
$Q^2 \gg 1\,GeV^2$ and $\nu \gsim  R_d \,\delta m_V^2/2 \approx 10 \,GeV$. 
The latter requirement ensures that the formation time of the 
vector meson exceeds the deuteron size, $\tau_f >  R_d$. 

Furthermore  Fig.1 demonstrates that  in the domain of 
vector meson dominance ($Q^2 = Q_0^2 \lsim 1\,GeV^2$)  
double scattering is by far dominant for $-t > 0.5\,GeV^2$: 
$\frac{(d\sigma_d(Q_0^2))/dt)_{full}}{(d\sigma_d(Q_0^2)/dt)_{Born}} 
\approx 5$.  
If final state interaction  vanishes at large $Q^2 = Q_1^2 \gg 1\,GeV^2$
and large energies $\nu \gsim 10\,GeV$, we expect 
the above ratio to approach unity.
Assuming that the $Q^2$-dependence of the initial ejectile 
production amplitude factorizes, and is approximately equal 
to the $Q^2$-dependence of the vector meson production cross 
section from free nucleons $d\sigma_N/dt$, we obtain: 
\begin{equation}
\frac{(d\sigma_d(Q_1^2)/dt)_{full}}{(d\sigma_d(Q_0^2)/dt)_{Born}}
\frac{(d\sigma_N(Q_0^2)/dt)}{(d\sigma_N(Q_1^2)/dt)} \longrightarrow 1.
\end{equation}
In both cases it is important to keep the coherence length constant 
or to account for its possible modification.   
     
\section{Summary and Outlook}
We discussed the coherent production of vector mesons from polarized 
deuterium as a tool to investigate the propagation of small size 
quark-gluon configurations, which are  initially produced 
in high energy lepton-nucleon interactions at 
large $Q^2$.  
A kinematic window was found where the differential cross section 
stems only from contributions of the final state interaction  
of the ejectile. 
Two methods for an investigation of color coherence effects,   
which are appropriate for future experiments at HERMES,  
were proposed. 
Note that
also the incoherent, diffractive production of vector mesons from 
polarized deuterium provides a variety of possibilities for  
detailed investigations of color coherence effects  
at HERMES \cite{FKMPSS}. 
Especially  detecting the recoil nucleon from deuteron 
break-up yields a possibility to explore the 
evolution of small size quark-gluon configurations at 
different, well defined length scales.

\end{document}